\documentclass[twocolumn,aps,showpacs,superscriptaddress,prmaterials,10pt]{revtex4-2}
\usepackage{graphicx}
\usepackage{color, soul}
\usepackage{amsmath}
\usepackage{natbib}
\usepackage{bm}
\usepackage{times}
\usepackage{url}
\usepackage{amssymb}
\usepackage{braket}
\usepackage{mathtools}
\usepackage{commath}
\usepackage{bigints}
\usepackage{booktabs}
\usepackage{multirow}
\usepackage{epstopdf}
\usepackage{pdfcomment}
\usepackage[normalem]{ulem}
\usepackage{mathrsfs}
\usepackage{xr}
\usepackage{cleveref}
\usepackage[top=30mm, bottom=25mm, left=25mm, right=30mm]{geometry}
\newcommand{\ti}[1]{\textit{#1}}

\usepackage{babel}
\usepackage[utf8]{inputenc}
\inputencoding{utf8}
\usepackage[mathscr]{euscript}
\usepackage{silence}
\definecolor{dgreen}{rgb}{0.0, 0.5, 0.0}

\newcommand\highlightReference[1]{%
  \expandafter\newcommand\csname highlightReference-#1\endcsname{}%
}

\begin{abstract}
While the effect of intrinsic defects on the electronic properties of half-Heusler compounds has been extensively discussed in literature, their effect on the lattice vibrations has received much less attention, due to the prohibitive computational demands.
This may lead to an erroneous description of the lattice thermal conductivity, which plays a crucial role in the thermoelectric efficiency, and for which there exists a significant discrepancy between ideal theoretical values and available experimental measurements.
In this article, we employ a combination of density-functional theory (DFT) and machine-learning interatomic potentials (MLIPs) to investigate how intrinsic defects affect the phonon spectra and lattice thermal conductivity of TaFeSb, alongside its electronic structure. 
The calculation of the formation energies of various defects identifies Fe interstitial atoms sitting at the vacant side of the HH crystal structure as the most likely to form, immediately followed by Sb substitution at Ta sites and by other antisite configurations. 
Phonon calculations illustrate that both defects generate localized phonon modes that significantly lower the lattice thermal conductivity, especially around room temperature. This reduction aligns the calculated values with available measurements, underscoring the critical role of intrinsic defects in reconciling the existing discrepancies between theory and experiment.
Our findings also reveal that these defects introduce localized electronic states, effectively reducing the theoretical electronic band gap and bringing it closer to the experimentally observed values. 
Finally, our analysis demonstrates the efficiency and effectiveness of machine-learning-based approaches to investigate defect-induced properties in complex materials for thermoelectric applications.
The proposed computational scheme may provide a feasible approach to incorporate the analysis of defects in the design of modern thermoelectric materials.
\end{abstract}

\begin{document}

\title{Understanding the role of defects in the lattice transport properties of half-Heusler compounds: a machine learning analysis}
%\title{Defect-Mediated Phononic and  Lattice Transport Properties in Half-Heuslers: A Machine Learning Analysis}

\date{\today}

\author{M. Yazdani-Kachoei}
\email{majidyazdani.kachoei@apctp.org}
\affiliation{Institute of Theoretical Physics
Jagiellonian University in Krakow ul. prof. Stanisława Łojasiewicza 11, PL-30-348 Kraków, Poland}
\affiliation{Institute of Physics, Nicolaus Copernicus University, 87-100 Toruń, Poland}
%\affiliation{Asia Pacific Center for Theoretical Physics, Pohang, 37673, Korea}
\author{B. Rabihavi}
\affiliation{Department of Physics, University of Tehran, Tehran, 14395-547, Iran}
\author{I. E. Brumboiu}
\affiliation{Institute of Physics, Nicolaus Copernicus University, 87-100 Toruń, Poland}
\author{S. Mehdi Vaez Allaei}
\email{smvaez@ut.ac.ir}
\affiliation{Department of Physics, University of Tehran, Tehran, 14395-547, Iran}
\author{I. {Di Marco}}
\email{igor.dimarco@physics.uu.se}
\affiliation{Institute of Physics, Nicolaus Copernicus University, 87-100 Toruń, Poland}
%\affiliation{Asia Pacific Center for Theoretical Physics, Pohang, 37673, Korea}
\affiliation{Department of Physics and Astronomy, Uppsala University, Box 516, SE-75120, Uppsala, Sweden}
%\affiliation{Department of Physics, POSTECH, Pohang, 37673, Korea}

\maketitle

%\begin{figure}
%\centering
%\includegraphics[width=0.48\textwidth]{band.eps}
%\end{figure}

\section{Introduction}
The fascinating properties of half-Heusler (HH) compounds have attracted significant attention due to their potential applications across various fields, including spintronics, optoelectronics, and solar cells, as indicated by numerous references in the literature~\cite{felser2015heusler, Sciadv_5_eaaw9337, Sciadv_3_e1602241,Nat.Commun_14_4722,Nat.Commun_7_11993,IOP_27_063001,Acsanm_5_569,PhysRevB.98.094410, PhysRevB.82.125210}. A field of particular interest for these materials is the production of thermoelectricity, which is expected to play a key role for the green-energy transition~\cite{Science_373_556, Sciadv_5_eaav5813, Nat.Commun_11_3142, Nat.Commun_10_270}. As a result, extensive work has been focused on the investigation of the properties of HH materials, both experimentally~\cite{Sciadv_5_eaaw9337, Sciadv_3_e1602241, Nat.Commun_14_4722, Nat.Commun_7_11993,Science_373_556, Sciadv_5_eaav5813, Nat.Commun_11_3142, Nat.Commun_11_3142, Nat.Commun_10_270} and theoretically~\cite{Acsanm_5_569, PhysRevB.98.094410, PhysRevB.82.125210, J.Alloys.Compd_828_154287, PhysRevMaterials.5.035407}. However, persistent discrepancies exist between theoretical predictions and experimental observations, which limits our ability to do large-scale materials engineering. For instance, NbCoSn, ZrNiSn and HfNiSn are predicted by density functional theory (DFT) to have better thermoelectric properties for p-type doping than for n-type doping, but the former regime cannot be realized in experiment~\cite{J.Appl.Phys._120_075104,APL.Materials_4_104804, J.Mater.Chem.A_8_14822,Appl.Phys.Lett._86_082105, ActaMaterialia_57_2757, Appl.Phys.Lett._79_4165,J.Appl.Phys._113_193705}.
%For instance, NbCoSn: while ab initio calculations within the density functional theory (DFT) framework predict NbCoSn as a p-type semiconductor\cite{J.Appl.Phys._120_075104},  experimental data\cite{APL.Materials_4_104804, J.Mater.Chem.A_8_14822} reveals its intrinsic n-type behavior.
%This discrepancy extends to ZrNiSn and HfNiSn, which experimentally exhibit n-type semiconductor characteristics\cite{Appl.Phys.Lett._86_082105, ActaMaterialia_57_2757, Appl.Phys.Lett._79_4165} but theoretical band structure calculations\cite{J.Appl.Phys._113_193705} classify them as p-type.
Quantitative disagreement on measurable properties is even more common. The measured total magnetic moment of AuMnSn~\cite{J.Alloys.Compd_274_136} amounts to 3.8 $\mu_B$, yet theoretical calculations~\cite{J.Phys.Chem.Solids_139_109328} predict the value of 4.02 $\mu_B$. More severe is the disagreement for the size of the band gap, which is crucial for obtaining accurate thermoelectric properties. Electronic structure calculations based on DFT give a band gap of 1.06 eV for TiCoSb~\cite{PhysRevB.86.045116} and 0.43 eV for TiNiSn ~\cite{PhysRevMaterials.5.035407},  in stark contrast to the corresponding experimental values of 0.57 eV~\cite{J.Appl.Phys._106_103703} and 0.12 eV~\cite{Zeitschrift_80_353}. Since the theoretical values are larger than the experimental ones, it is difficult to argue that this problem may be connected to the standard underestimation associated to local or semi-local exchange-correlation functionals~\cite{martin_book}.
In fact, electronic structure calculations of TaFeSb demonstrate that the estimate of the band gap even worsens if more advanced functionals are used~\cite{Nat.Commun_10_270}.

One plausible explanation for the disagreement between theory and experiment lies in the correspondence between the system that is modeled and the one that is probed. An issue of concern in experiment is the quality of the synthesized samples, which may contain a high density of defects for HH compounds. Recognizing the significance of intrinsic point defects can greatly enhance the agreement between theoretical predictions and experimental measurements for magnetic, electronic and vibrational properties.  For example, theoretical calculations~\cite{J.Phys.Chem.Solids_139_109328} confirm that doping AuMnSn with Au atoms at the Mn site reduces the predicted magnetic moment of AuMnSn to 3.75 $\mu_B$, which aligns closely with the aforementioned experimental measurements. Similarly, considering point defects can improve the theoretical estimate of the electronic band gap of TiCoSb, reducing it to 0.62 eV, in a much better agreement with experimental observations~\cite{PhysRevMaterials.5.035407}. Similar improvements are reported for the electronic properties of other HH compounds, such as ZrNiSn and TiNiSn, where point defects play a crucial role in matching theory and experiment~\cite{PhysRevMaterials.5.035407}.
Finally, defect engineering in HH compounds can also be used to create systems of XY$_{1+ \delta}$Z chemical formula that are intermediate between HH and full-Heusler (FH). Examples include TiFe$_{1.33}$Sb~\cite{10.1039/C7DT03787B}, ZrNi$_{1+\delta}$Sn ($\delta = 0.00 - 0.05$)~\cite{0.1038/natrevmats.2016.32}, TiRu$_{1+\delta}$Sb ($\delta = 0.15 - 1.00$)~\cite{10.1038/s41467-021-27795-3}, and MCo$_{1.5}$Sn (M = Ti,
Zr, Hf)~\cite{10.1016/j.mtphys.2020.100251}. The value of $\delta$ offers another path to modulate the physical properties of these compounds, and in particular their thermoelectric efficiency.

The role of defects in HH compounds suitable for thermoelectric applications has been already addressed in previous studies~\cite{J.Phys.Chem.Solids_139_109328, PhysRevMaterials.5.035407, RevModPhys.86.253}. However, these works are mostly focused on the electronic properties, while it is well established that lattice properties, and in particular the lattice thermal conductivity ($\kappa_L$), are a key feature to obtain a high figure of merit~\cite{OJHA2024107996}. 
Previous literature has emphasized that theory tends to overestimate the lattice thermal conductivity of HH compounds substantially, as shown by works on TaFeSb, ZrNiSn, and HfNiSn~\cite{Nat.Commun_10_270,PhysRevMaterials.7.104602,Comp.Mat.Sci.202.110938,PhysRevB.77.184203, 10.1088/0953-8984/11/7/004,PhysRevX.4.011019,10.1007/s11664-009-1032-8,PhysRevX.4.011019}. 
%In this regard, large discrepancies are reported in literature. Our previous calculations~\cite{PhysRevMaterials.7.104602} and other studies~\cite{Comp.Mat.Sci.202.110938} predict significantly higher $\kappa_L$ for pure TaFeSb, especially around room temperature, compared to experimental values~\cite{Nat.Commun_10_270}. Experimental measurements report $\kappa_L$ values of 7.5 W/mK for ZrNiSn and 8.8 W/mK for HfNiSn~\cite{PhysRevB.77.184203, 10.1088/0953-8984/11/7/004}, while theoretical predictions suggest much higher values of 17.9 W/mK and 17.5 W/mK, respectively~\cite{PhysRevX.4.011019}. For HfNiSn, the experimental and theoretical results show a significant disparity, with values of 4.8 W/mK~\cite{10.1007/s11664-009-1032-8} and 19.5 W/mK~\cite{PhysRevX.4.011019}, respectively.
Understanding how defects affect phonons and lattice thermal conductivity has become even more prominent with the rise of 19-electron HH compounds, where defects are necessary to stabilize the material below the convex hull~\cite{acs.chemmater.6b04583}. This has motivated the work by Ren \textit{et al.}~\cite{10.1038/s41467-023-40492-7}, who have calculated the phonon dispersion in the perfectly stoichiometric ZrNiBi as well as in the defective Zr$_{0.75}$NiBi. Zr vacancies were shown to induce the formation of flat optical-phonon bands leading to a reduced group velocity and thus a smaller lattice thermal conductivity with respect to the pure sample. 
Despite this seminal work, there is a severe lack of studies regarding the role of defects on the lattice transport properties in HH compounds. This shortage is due to the computational challenges involved in the calculation of the first-principles phonon spectra that are needed to obtain $\kappa_L$ through the Boltzmann Transport Equation (BTE)~\cite{Comput.Phys.Commun.185.1747}. Since modeling basic defects in DFT requires the usage of large supercells, standard computational approaches for phonons, as e.g. density-functional perturbation theory (DFPT)~\cite{PhysRevB.43.7231,PhysRevB.55.10355} or the finite-displacement method~\cite{togo_2015_first}, become extremely demanding. 
While classical molecular dynamics (MD) offers a faster alternative, it lacks a sufficient accuracy to assure a quantitative comparison with experiment, as commonly used potentials like Lennard-Jones fail to properly capture the complex interatomic interactions in HH compounds. Recent studies~\cite{10.1016/j.cpc.2020.107583, 10.1016/j.commatsci.2019.03.049, 10.1038/s41524-024-01390-8} have suggested that these shortcomings of MD may be remedied by the usage of machine-learning interatomic potentials (MLIPs), which retain an accuracy comparable to first-principle techniques while offering a much better scaling with system size. This becomes a key factor for the description of defects, especially if one intends to account for higher order terms  in the lattice vibrations (anharmonic force constants).

In this article, we employ a combination of DFT and MLIPs to clarify how point defects influence not only the electronic structure but also the phonon characteristics and thermal conductivity of HH compounds. TaFeSb is chosen as a representative test case, due to its astonishingly high thermoelectric efficiency  as well as the existence of a substantial discrepancy between theory and experiment~\cite{Zhu.NatCommun.10.270}. 
This disagreement manifests first for the electronic band gap, which is estimated to be around 0.53 eV from infrared spectroscopy~\cite{Zhu.NatCommun.10.270}. The theoretical predictions from electronic structure calculations provide values in the range of 0.84-0.87 eV in the generalized gradient approximation (GGA) of Perdew, Burke, and Ernzerhof (PBE)~\cite{J.phys.Mat.2.035002,Grytsiv.Intermetallics.111.106468,PhysRevMaterials.7.104602} up to 0.9 eV for the modified Becke–Johnson (mBJ) potential of Tran and Blaha~\cite{Zhu.NatCommun.10.270}.
%Similarly, alternative theoretical calculations based on the PBE-GGA approach converged around values of 0.86 eV~\cite{J.phys.Mat.2.035002} and 0.87 eV~\cite{Grytsiv.Intermetallics.111.106468}. Our prior computations~\cite{PhysRevMaterials.7.104602} also aligned with this trend, suggesting a band gap of 0.84 eV for this compound.
Similarly, experimental measurements of the lattice thermal conductivity provide values of 9 {W}{m$^{-1}$K$^{-1}$} at room temperature, while theoretical predictions range from 18 to 30 {W}{m$^{-1}$K$^{-1}$}~\cite{PhysRevMaterials.7.104602,Comp.Mat.Sci.202.110938}. 

The text is structured as follows. After this introduction, methodological details are illustrated in Section~\ref{cal-sec}. The energetic stability of defects and their effect on the electronic excitation spectra are presented in Section~\ref{form-sec} and Section~\ref{elec-sec}, respectively. The phonon dispersions are analyzed in Section~\ref{phon-sec}, while the lattice thermal conductivity is discussed in Section~\ref{cond-sec}. Finally, the conclusions of our study and an outlook for future research are illustrated in Section~\ref{conc-sec}.
%endeavor involves the computation of formation energy values for the compound in the presence of various defects and an in-depth exploration of the electronic properties of the most probable defects, identified by their lowest formation energies. Additionally, as a key strength of this paper, we investigate the phonon properties and $\kappa_L$ in the presence of those defects.

\section{Methodological details}\label{cal-sec}
HH compounds with the chemical formula XYZ crystallize in the space group $F43m$ (No. 216), adopting a cubic structure consisting of three interpenetrating face-centered cubic (fcc) sublattices~\cite{MRS.Bulletin_47_555}. The constituent atoms X, Y and Z occupy the Wyckoff positions (0, 0, 0), (0.25, 0.25, 0.25), and (0.5, 0, 0), respectively~\cite{MRS.Bulletin_47_555}. To simulate defect formation, a $3 \times 3 \times 3$ supercell is constructed, corresponding to a concentration of about 3.7\%. Previous studies~\cite{Intermetallics_46_103} have shown that this concentration is already representative of the dilute limit and larger supercells do not significantly impact the value of the formation energies. All electronic structure calculations are performed through DFT~\cite{Kohn1965, Hohen1964} using the Quantum Espresso package~\cite{QE_1, QE_2}, based on the projector augmented wave (PAW) method~\cite{PhysRevB.50.17953}. The exchange-correlation functional is treated in GGA, parametrized according to PBE~\cite{PBE-GGA}. A kinetic energy cutoff of 100 Ry is chosen for wavefunctions, while a value of 1000 Ry is used for charge density and potential. The convergence thresholds for total energy and Kohn-Sham potential are set at $10^{-6}$ Ry and $10^{-10}$, respectively. Additionally, the convergence thresholds for ionic forces and stress are specified as $10^{-4}$ Ry/a.u. and 0.50 Kbar. The sampling of the Brillouin zone is performed with a Monkhorst–Pack grid \cite{mon76} of $6 \times 6 \times 6$ \textbf{k}-points. The electronic band structure of the supercell are unfolded to the unit cell by means of the BandUP code~\cite{PhysRevB.89.041407, PhysRevB.91.041116}.
%### MTP ####

As mentioned in the introduction, interatomic interactions necessary for the lattice vibrations are described by means of a class of MLIPs, namely the moment-tensor potentials (MTPs)~\cite{shapeev_2016_moment}. The parameters of MTPs are optimized on training sets created by means of ab-initio molecular dynamics (AIMD) simulations, which are described below. MTPs are multi-component potentials obtained from a model that expresses the total energy of a system as a summation of contributions \(V\) from the neighborhoods \(n_i\) defined by the local atomic environments surrounding each atom up to a given cut-off radius, which in our case is set to be of 5~{\AA}. The atomic environment, or neighborhood \(n_i\), of the \(i\)-th atom includes its atomic type (\(z_i\)), the atomic types of its neighboring atoms (\(z_j\)), and finally their relative positions (\(r_{ij}\)). Thus, the total energy can be expressed as~\cite{shapeev_2016_moment}:
\begin{equation}
E \equiv E^{\text{MTP}} = \sum_{i=1}^{N} V(n_i) \tag{1}
\end{equation}
To train the MTP model, we need to minimize the following objective (cost) function:
\begin{align}
\sum_{k=1}^K & \left[ w_e(E_{k}^{\text{AIMD}} - E_{k}^{\text{MTP}})^2 + w_f \sum_{i=1}^{N} |f_{k,i}^{\text{AIMD}} - f_{k,i}^{\text{MTP}}|^2 \right. \nonumber \\
& \left. + w_s \sum_{i,j=1}^{3} |\sigma_{k,ij}^{\text{AIMD}} - \sigma_{k,ij}^{\text{MTP}}|^2 \right] \rightarrow \min \tag{2}
\end{align}
In this formulation, \(E_{k}^{\text{AIMD}}\), \(f_{k,i}^{\text{AIMD}}\), and \(\sigma_{k,ij}^{\text{AIMD}}\) represent the energy, atomic forces, and stresses from the training set, while \(E_{k}^{\text{MTP}}\), \(f_{k,i}^{\text{MTP}}\), and \(\sigma_{k,ij}^{\text{MTP}}\) are the corresponding values calculated using the MTP. Here, \(K\) denotes the number of configurations in the training set, and \(w_e\), \(w_f\), and \(w_s\) are non-negative weights assigned to energies, forces, and stresses in the optimization process, respectively. 
In our study, these weights are set to 1, 0.1, and 0.1, which are similar to those obtained by optimizing the accuracy of diffusion processes in bulk materials~\cite{10.1016/j.commatsci.2019.03.049} and anharmonic force constants in two-dimensional materials~\cite{10.1016/j.cpc.2020.107583}.
The training set comprises approximately 3000 configurations created via AIMD and solved via DFT. Most configurations are generated for $2 \times 2 \times 2$ supercells of pristine TaFeSb, using a mesh of $2 \times 2 \times 2$ \textbf{k}-points.
Additional configurations are generated for $2 \times 2 \times 2$ supercells of defective TaFeSb, including compositions such as TaFe\(_{1.125}\)Sb or Ta\(_{0.875}\)FeSb\(_{1.125}\), and using a mesh of {$2 \times 2 \times 2$} \textbf{k}-points.
AIMD simulations were performed by means of the LAMMPS package~\cite{thompson_2022_lammps}, using the isothermal-isobaric (NPT) equilibration with temperatures between 300~K and 1100~K, at zero pressure. In MD, for training, 50000 steps are used, with a time step of 1 fs, whereas the box size is the same as the size of the supercells used for training and force constant calculations. 

%3x3x3 supercell calculation of formation E, compared with bulk phases
%select most probable defects -> get DOS, bands of supercell -> unfold to unit cell

% pure case:
% DFT+DFPT (QE) -> harmonic F.C. -> phonons (with QE directly)
% DFT+FD (QE) ->  harmonic F.C. -> phonons (PHONOPY)
% MPT+FD (MLIP) -> harmonic F.C. -> phonons (PHONOPY)
% DFT+DFPT (QE+D3Q) -> harmonic + anharmonic F.C. -> TCOND (KALDO)
% MPT+MD (MLIP+LAAMPS) -> harmonic + anharmonic F.C. -> TCOND (KALDO)

% defects:
% MPT+FD (MLIP) -> harmonic F.C. -> phonons (PHONOPY)
% MPT+MD (MLIP+LAAMPS) -> harmonic + anharmonic F.C. -> TCOND (KALDO)

After the training of the MTPs, the finite displacement (FD) method is used to calculate the phonon spectra with and without defects, by means of the PHONOPY code~\cite{togo_2015_first}. To assess the errors associated to the usage of MLIPs, the phonon dispersion of pristine TaFeSb is also calculated directly via DFT+FD, again by means of PHONOPY, as well as via DFT+DFPT, using the implementation of Quantum Espresso~\cite{PhysRevB.43.7231,PhysRevB.55.10355}. 
Finally, the lattice thermal conductivity $\kappa_L$ is calculated using the BTE in the relaxation-time approximation (RTA), as implemented in the KALDO package~\cite{barbalinardo_2020_efficient}. The second (harmonic) and third order (anharmonic) force constants, which are needed as input, are calculated through MD simulations based on the trained MTPs~\cite{novikov_2019_improving}, by means of the LAMMPS package~\cite{thompson_2022_lammps}. For sake of comparison, the thermal conductivity of pristine TaFeSb is also calculated using second and third order force constants extracted from DFT+DFTP calculations by Quantum Espresso thanks to the D3Q code~\cite{PhysRevB.87.214303,PhysRevB.88.045430, PhysRevB.91.054304,PhysRevB.101.205419}.

\section{Formation Energy}\label{form-sec}
\begin{figure}[t]
\centering
\includegraphics[width=0.48\textwidth]{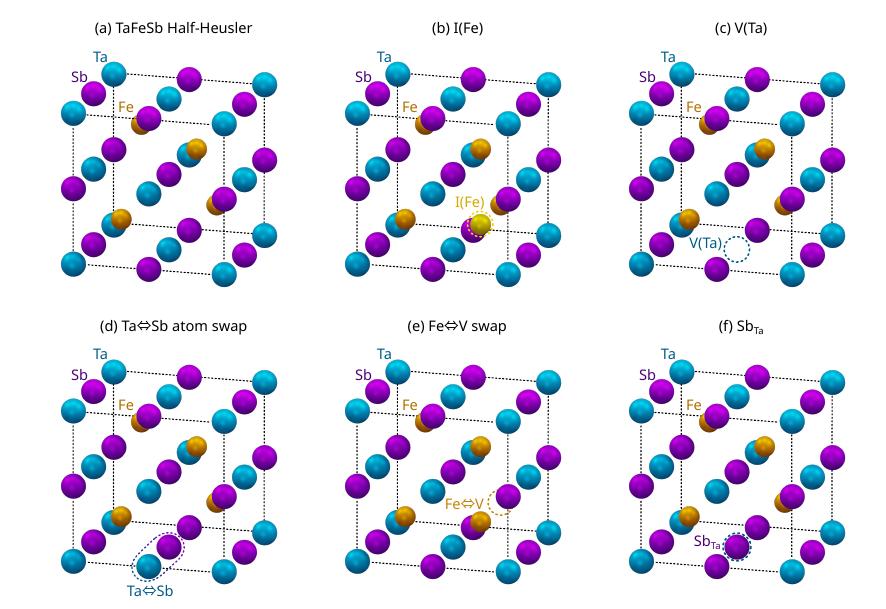}
\caption{Illustration of all the defects considered in this work, depicted in the conventional unit cell: (a) pristine HH structure; (b) Fe interstitial impurity; (c) a vacancy in the Ta sublattice; (d) a swap of Ta and Sb atoms; (e) a swap of an atom of Fe and a vacancy in the interstitial lattice; (f) an antisite defect $\text{Sb}_\text{Ta}$. Calculations are performed on $3\times3\times3$ supercells, approaching the dilute limit.}
\label{fig:defects}
\end{figure}

HH compounds differ from FH compounds for having one less fcc sublattice in the crystal structure. In the unit cell, this is reflected by a vacancy at $(0.75, 0.75, 0.75)$~\cite{MRS.Bulletin_47_555}.
In the study by Grytsiv \ti{et al.}, electronic structure calculations of TaFe$_{1+x}$Sb, for $x$ comprised between 0 and 1, revealed the tendency to have Fe impurities occupying sites in the vacant sublattice, which results in a partially filled impurity band~\cite{Grytsiv.Intermetallics.111.106468}.
%However, other defects may occur which have not been investigated yet. 
Moreover, Tavassoli \ti{et al.} showed that, for TiFeSb, the single-phase region of the Heusler phase is significantly shifted from the perfect stoichiometry to an Fe-rich composition, also due to Fe/Ti substitution~\cite{10.1039/C7DT03787B}.
Likewise, Chibueze \ti{et al.} found a small, negative formation energy for Au$\Rightarrow$Mn antisites (Au atom in the Mn position), Sn$\Rightarrow$Mn antisites, and Au interstitial defects, suggesting that they are likely to form spontaneously during the crystal growth process~\cite{J.Phys.Chem.Solids_139_109328}.
Motivated by this wealth of works, as well as by the aforementioned discrepancy for band gap and lattice thermal conductivity, we proceed to investigate four types of intrinsic defects (see Figure~\ref{fig:defects} for an illustration): 1) vacancies $\text{V}_\alpha$ (where the atom $\alpha$ is removed from a site); 2) antisite defects $\beta_\alpha$ (where the atom $\alpha$ of a given site is replaced by the atom $\beta$ of a different type); 3) interstitial defects $\text{I}_\alpha$ (where the atom $\alpha$ is added to the vacant 4d site of the $C1b$ crystal structure); 4) atom swaps $\beta \Leftrightarrow \alpha$ (where the atoms $\alpha$ and $\beta$ at different sites are swapped). Notice that only the last type of defects preserve the precise stoichiometry of TaFeSb, while the other ones lead to a change of atomic content. The stability of these defects is estimated through the following formation energies $\Delta E$:
\begin{center}
\begin{equation}
\Delta E(\text{V}_\alpha) = E^\text{defect} - E^\text{perfect} + \mu_\alpha ,
\end{equation}
\begin{equation}
\Delta E(\beta_\alpha) = E^\text{defect} - E^\text{perfect} + \mu_\alpha - \mu_\beta ,
\end{equation}
\begin{equation}
\Delta E(\text{I}_\alpha) = E^\text{defect} - E^\text{perfect} -\mu_\alpha ,
\end{equation}
\begin{equation}
\Delta E(\beta \Leftrightarrow \alpha) = E^\text{defect} - E^\text{perfect} .
\end{equation}
\end{center}
%where $\Delta E(I_\alpha)$, $\Delta E(V_\alpha)$, $\Delta E(\beta_\alpha)$, and $\Delta E(\beta \Leftrightarrow \alpha)$ represent the formation energies for interstitial defect, vacancy, substitution, and atom swapping, respectively. 
\begin{figure}[b]
\centering
\includegraphics[width=0.48\textwidth]{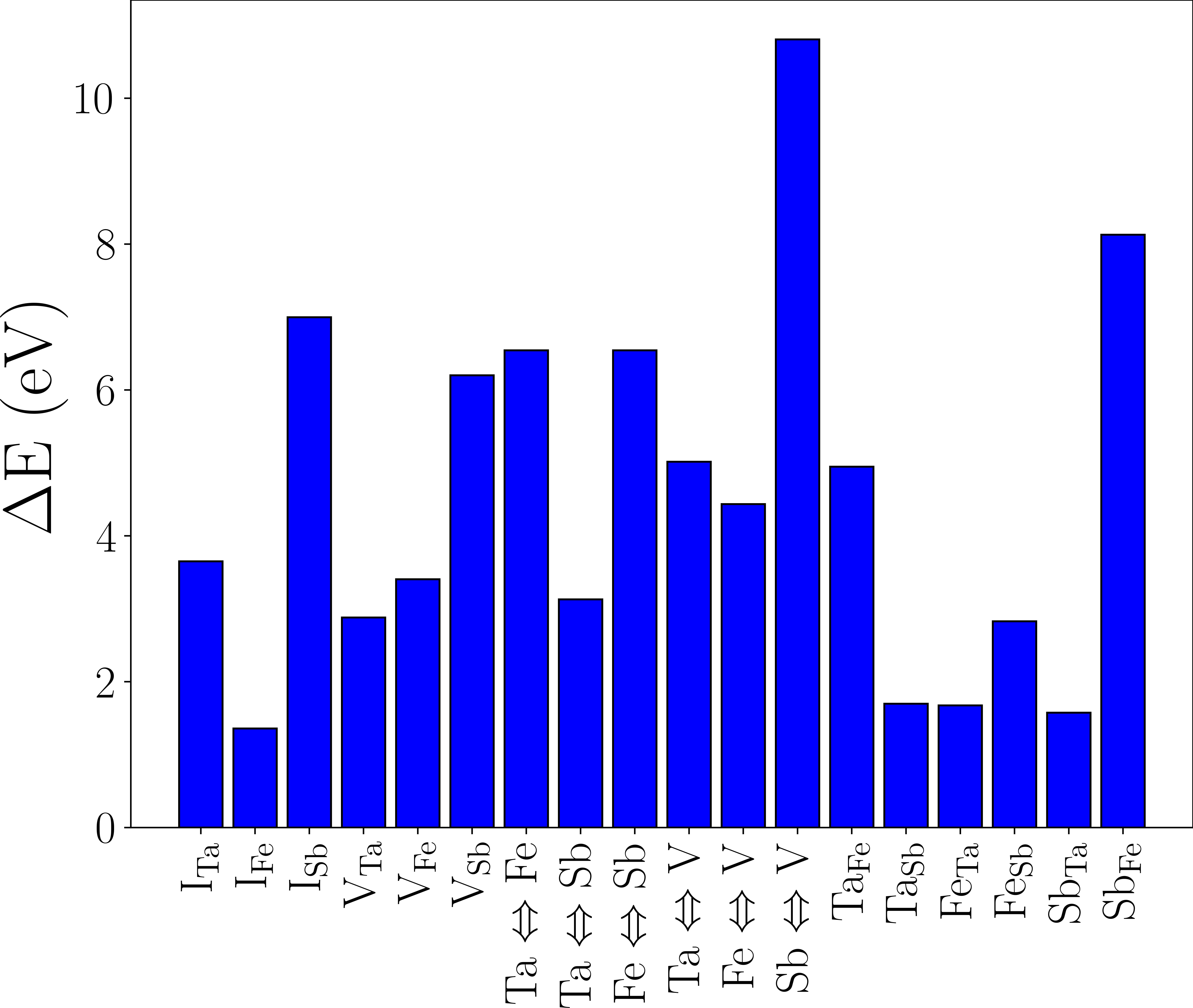}
\caption{Formation energy of selected defects in TaFeSb, including vacancies $\text{V}_\alpha$, antisite defects $\beta_\alpha$, interstitial defects $\text{I}_\alpha$, and atom swaps $\beta \Leftrightarrow \alpha$. Calculations are performed on $3\times3\times3$ supercells, approaching the dilute limit.}
\label{fig:delta_e}
\end{figure}
In the expressions above, $E^\text{defect}$ and $E^\text{perfect}$ represent the total energies of the supercells with and without the defect, respectively. The chemical potentials $\mu_\alpha$ and $\mu_\beta$ for atoms $\alpha$ and  $\beta$ are calculated by using the ground-state energies per atom of the corresponding elemental solids. Thus, the body-centered cubic (bcc) structure was used for Ta and Fe, while a trigonal structure ($R\bar{3}m$) was used for Sb, as determined through the Materials Project database~\cite{Jain2013}.
The calculated formation energies displayed in Figure~\ref{fig:delta_e} indicate that the Fe interstitial defect ($\text{I}_\text{Fe}$) has the lowest formation energy, amounting to about 1.28~eV. The formation energy of $\text{Sb}_\text{Ta}$ is approximately 1.50~eV, which is slightly larger than that of $\text{I}_\text{Fe}$. At a slightly higher energy, we can find $\text{Ta}_\text{Sb}$ and $\text{Fe}_\text{Ta}$. All the other defects have substantially larger formation energies.
These values are in line with the results obtained for CoZrBi, where Co interstitial defects are predicted to be the most favorable ones, with a formation energy of 1.02 eV~\cite{PhysRevB.94.075203}, as well as for TiNiSn, where interstitial defects of Ni are are predicted to be the most favorable ones, with a formation enthalpy of 0.77 eV~\cite{Intermetallics_46_103}. Our results are also in line with the a comprehensive study on NbCoSn, TiCoSb, ZrNiSn, and TiNiSn, albeit the authors used a different convention for the sign of the formation energy~\cite{PhysRevMaterials.5.035407}.
Having identified the most favorable defects that will arise during growth, in the next section we will investigate how they affect the electronic structure, phonon properties, and lattice thermal conductivity of TaFeSb. For simplicity, we will focus exclusively on Fe interstitial defects  $\text{I}_\text{Fe}$ and $\text{Sb}_\text{Ta}$ antisite defects.

\section{Electronic structure}\label{elec-sec}
\begin{figure}
\centering
\includegraphics[width=0.48\textwidth]{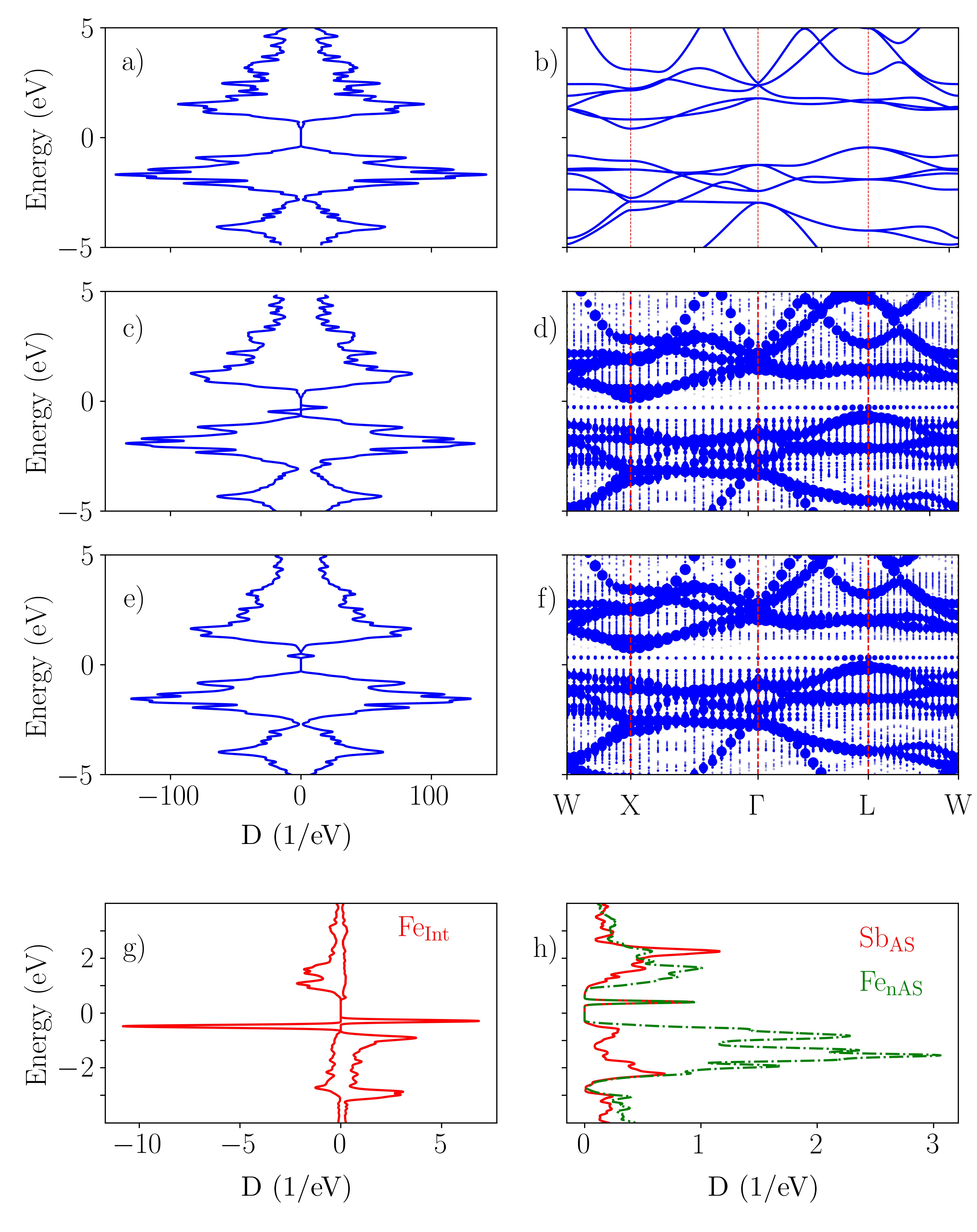}
\caption{Total DOS for TaFeSb without defects (a), with Fe interstitial defects at 3.7\% (c) and with $\text{Sb}_\text{Ta}$ antisite defects at 3.7\% (e). Corresponding band structures along high-symmetry paths in the Brillouin zone for TaFeSb are also shown, in panels (b), (d), and (f), respectively.
In panel (g), the PDOS of the interstitial Fe atom for TaFeSb with Fe interstitial defects at 3.7\% is also shown, in red, for a smaller energy range. In panel (h), for TaFeSb with $\text{Sb}_\text{Ta}$ antisite defects at 3.7\%, the PDOS of the Sb atom at the Ta site is shown, in red, together with the average PDOS of the Fe atoms around it, in green.}
\label{fig:band_dos}
\end{figure}
%For the TaRuAs compound, the $As_{Ta}$ antisite defect exhibits the lowest formation energy, measuring approximately 0.95 eV.
The electronic structure and the density of states (DOS) of TaFeSb without defects as well as for $\text{I}_\text{Fe}$ and $\text{Sb}_\text{Ta}$ defects are displayed in Figure~\ref{fig:band_dos}. 
With respect to perfect TaFeSb, the interstitial defect $\text{I}_\text{Fe}$ introduces a narrow impurity band within the band gap, just below the Fermi level. The projected density of states (PDOS) of the interstitial Fe atom demonstrates that those states represent the main contribution to the impurity band, compare panels (c) and (g) of Figure~\ref{fig:band_dos}. The results for TaFeSb with the $\text{Sb}_\text{Ta}$ defect also generate a narrow impurity band, which is however located above the Fermi level. 
The analysis of the PDOS shows that this impurity band originates from states that are more delocalized, including both the replaced Sb atom as well as the Fe atoms in its neighborhood, see panel (h) of Figure~\ref{fig:band_dos}.
The formation of these detached impurity bands can affect transport, and in particular the transmission function, which in turn affects the thermoelectric properties~\cite{PhysRevB.94.115414, Rezaei_2021}.
Interestingly, the band gap of perfect TaFeSb is reduced in both cases, going from  0.85~eV to 0.40~eV and 0.80~eV for $\text{I}_\text{Fe}$ and $\text{Sb}_\text{Ta}$, respectively. These values are much closer to the experimentally reported value from infrared spectroscopy~\cite{Zhu.NatCommun.10.270}. 
The impurity-derived states and the reduction of the band gap observed in Figure~\ref{fig:band_dos} are expected to increase the electronic thermal conductivity as well as to reduce the Seebeck coefficient, thus affecting the thermoelectric properties. The precise quantitative behavior depends on the position of the Fermi energy, and thus on the carrier concentrations, as emphasized in previous studies~\cite{PhysRevB.94.075203}. Since we have already investigated the connection between carrier concentration and thermoelectric properties for TaFeSb in a previous work~\cite{PhysRevMaterials.7.104602}, here we focus on the reduction of the lattice thermal conductivity, which is expected to play a more relevant role for solving the existing discrepancies between theory and experiment.

%\textcolor{red}{The efficiency of a thermoelectric power generator can be explained via the following formula:
%\begin{equation}
%   \eta = \frac{VI}{\dot{Q}_h},
%\end{equation}
%where $V$ is the bias voltage, $\dot{Q}_h$ is the heat rate lost from the hot lead, and $I$ is the electric current. The electric current can be calculated using Landauer-B\text{$\Ddot{u}$}ttiker formalism~\cite{Datta_1995}:
%\begin{equation}
%    I = \frac{2|e|}{h} \int T(E)f_0(E)(f_h-f_c]E
%\end{equation}
%where, $h, ~e, ~T$ are the Planck constant, electron charge,  the total transmission function. The $f_h$ and $f_c$ Fermi–Dirac distribution of the hot and cold sides of a thermoelectric system. T(E) can be written as:~\cite{10.1063/1.4962346}
%\begin{equation}
%    T(E) = Tr[\Gamma_LG_d\Gamma_RG_d^+],
%\end{equation}
%in which $Tr$ is the trace operator, $\Gamma_{L,R}$  are the broadening factors, and superscript‘+’ shows conjugate transpose. Moreover, the DOS can be explained using the following equation:~\cite{10.1063/1.4962346}
%\begin{equation}
%    DOS(E) = \frac{1}{2 \pi} [G_d(\Gamma_L + \Gamma_R)G_d^+]
%\end{equation}
%}

% defects -> electrons                     -> electron thermal conductivity -> speculation on ZT
% defects -> electrons                     -> Seebeck coeffcient -> speculation on ZT
% defects -> phonons -> lattice thermal conductivity -> speculation on ZT

\section{Phonon properties}\label{phon-sec}
\begin{figure}
\centering
\includegraphics[width=0.48\textwidth]{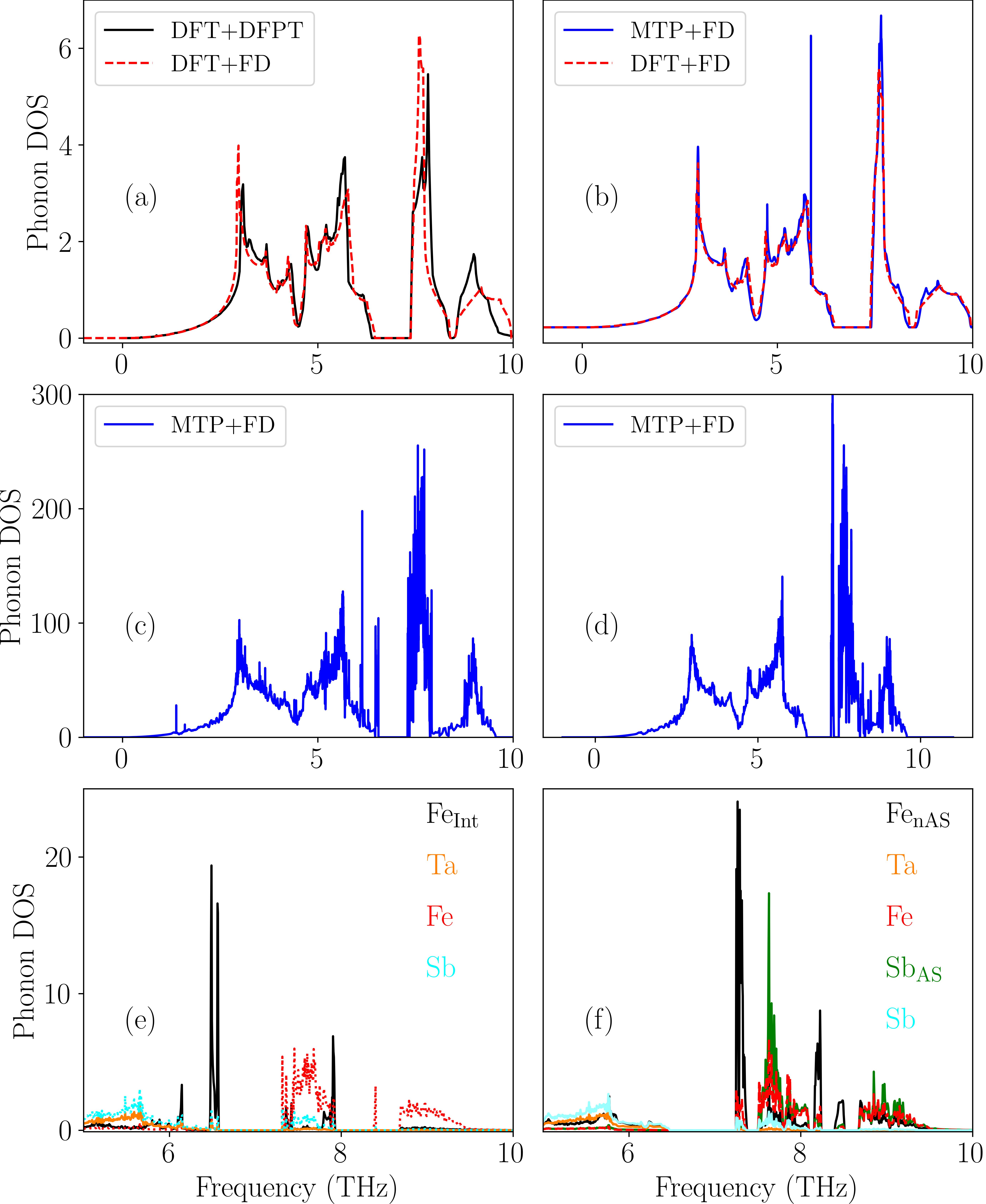}
\caption{The phonon DOS calculated for perfect TaFeSb using different computational approaches is shown in panels (a) and (b). The phonon DOS calculated by means of MTP+FD for TaFeSb with Fe interstitial defects at 3.7\% is shown in panel (c). The corresponding phonon PDOS of the component atoms is shown in panel (e), including the contribution of the Fe interstitial atom in comparison to the average contribution of the other atoms in the supercell. The phonon DOS calculated by means of MTP+FD for TaFeSb with $\text{Sb}_\text{Ta}$ antisite at at 3.7\% is shown in panel (d). The corresponding phonon PDOS of the component atoms is shown in panel (f), including the contribution of the additional Sb from the $\text{Sb}_\text{Ta}$ antisite defect, alongside the average contribution from the Fe atoms in its neighborhood as well as the average contributions from the other atoms in the supercell.}
\label{fig:pdos}
\end{figure}

 Figure~\ref{fig:pdos} shows the phonon DOS of TaFeSb without defects as well as for $\text{I}_\text{Fe}$ and $\text{Sb}_\text{Ta}$ defects. The results for perfect TaFeSb are shown in panels (a) and (b), where three sets of data are reported, as obtained via DFT+DFPT, DFT+FD and via MTP+FD. Comparing the three curves reveals a very good agreement for a wide range of energies, with some differences appearing only above 5~THz. The excellent agreement of the curves in panel (b) sets the range of accuracy of the MTPs achieved during the training.  
The phonon DOS for TaFe$_{1.037}$Sb with interstitial Fe is shown in Figure~\ref{fig:pdos}(c), while the one for Ta$_{0.963}$Sb$_{1.037}$ with $\text{Sb}_\text{Ta}$ antisite is shown in Figure~\ref{fig:pdos}(d).
For perfect TaFeSb, a gap is visible at around 7~THz, whose size amounts to 0.95~THz.
The additional Fe at the interstitial position decreases the size of this gap to 0.68~THz, creating some modes at frequencies around 6.5~ THz. The additional Fe also pushes down the lowest optical modes near 7.5~THz, albeit slightly.
The presence of antisite $\text{Sb}_\text{Ta}$ defects also decreases the gap between acoustic and optical modes, to 0.74~THz. However, in contrast to $\text{I}_\text{Fe}$, it creates only higher energy optical modes, at frequencies above 7.25~THz.
To better understand the origin of these additional bands, the partial phonon DOS of the defected cases is displayed in Figure~\ref{fig:pdos}(e) and Figure~\ref{fig:pdos}(f). For $\text{I}_\text{Fe}$, the additional phonon modes have a pure Fe-interstitial character, while the contribution of the other atoms is negligible. For $\text{Sb}_\text{Ta}$ antisite defects, the additional phonon modes are mainly localized at the Fe atoms located in the nearest neighborhood of the substituted Sb atom, with a minor contribution arising from the second shell of Fe neighbors. The contribution of all the other atoms, including the substituted Sb, to this narrow phonon band is negligible. Instead, the substituted Sb atom contributes to the optical branch above the gap.

\section{Lattice thermal conductivity}\label{cond-sec}
\begin{figure}
\centering
\includegraphics[width=0.48\textwidth]{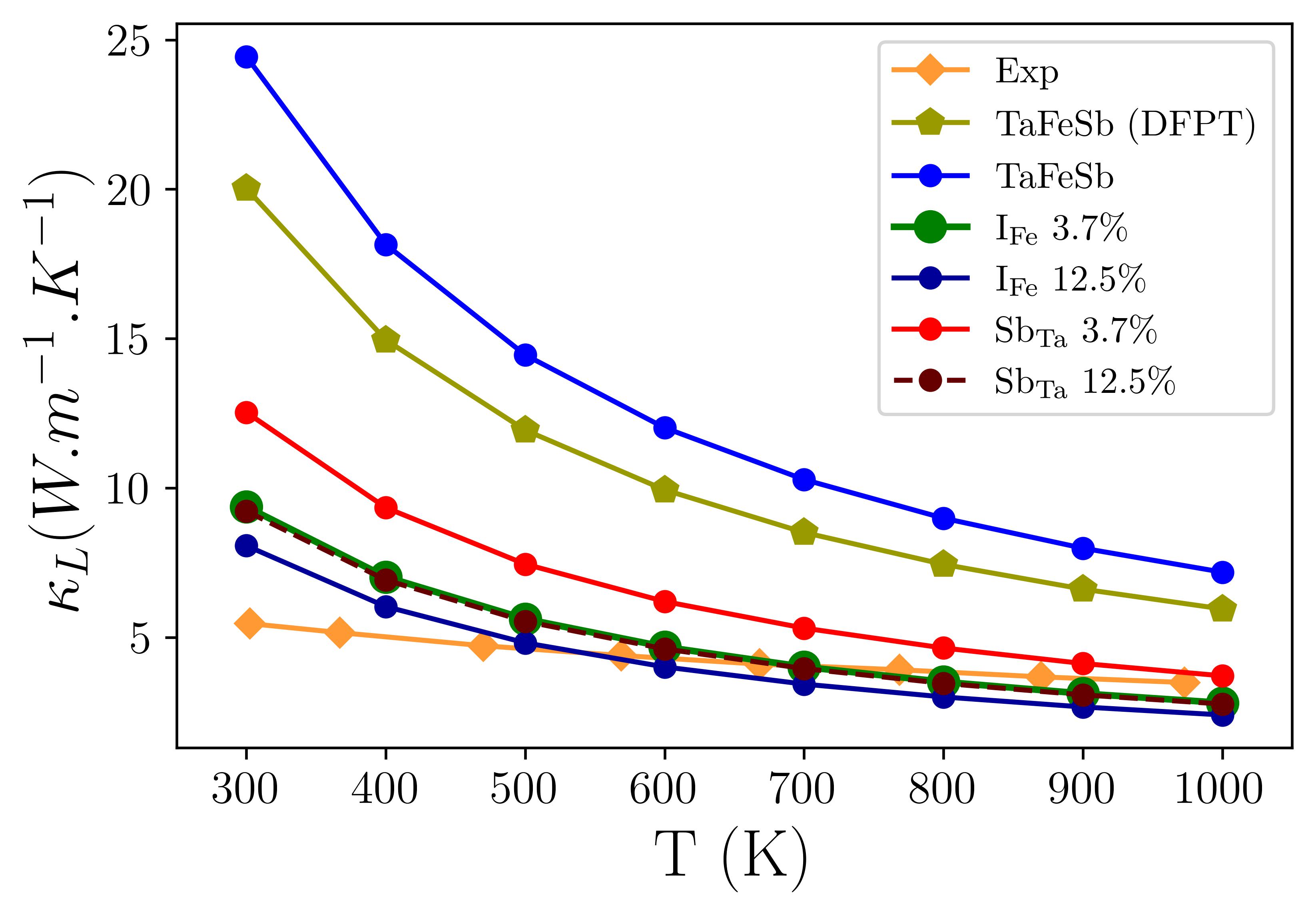}
\caption{Calculated lattice thermal conductivity ($\kappa_L$) of pure and defected TaFeSb using various approaches along with experimental results~\cite{Zhu.NatCommun.10.270}. 
%The calculations using the Slack approach (TaFeSb-Slack) are done in our previous work~\cite{PhysRevMaterials.7.104602}
All theoretical curves but one are calculated by means of the KALDO package, using second and third order force constants obtained from MTP+MD simulations. For pristine TaFeSb, an additional curve is shown (blue circles) evaluated with second second and third order force constants obtained from DFT+DFTP.}.\label{fig:k_lat}
\end{figure}

The introduction of point defects through doping and alloying has historically been a paramount and steadfast approach to enhance the thermoelectric efficiency of materials. In this context, intrinsic defects assume a pivotal role by exerting influence over carrier concentration and, more critically, impacting phonon properties, which, in turn, culminates in a substantial reduction of the lattice thermal conductivity - a key facet for enhancing thermoelectric performance~\cite{10.1007/s11664-012-2048-z, 10.1039/D1CS00347J, 10.1039/D0EE03014G}.
For instance, the relaxation time of point defects is inversely proportional to the square of the phonon frequency, which significantly affects the scattering of high-frequency phonons~\cite{10.1039/D1CS00347J}.  Additionally, phonon-vacancy scattering can greatly hinder the propagation of heat-carrying phonons, leading to a reduction in lattice thermal conductivity~\cite{10.1039/D1CS00347J}. These expectations are confirmed by the inspection of Figure~\ref{fig:k_lat}, where the calculated lattice thermal conductivity is reported alongside the experimental data~\cite{Nat.Commun_10_270}. As a term of comparison, we first look at the data for perfect TaFeSb, as calculated from second and third order force constants obtained via DFT+DFTP and via MTP+MD.
Two main features emerge. First, the theoretical curves overestimate the experimental data by a factor around 4 for the most accurate DFT+DFTP approach. Second, the less accurate MTP+MD approach worsens the results, by increasing the predicted conductivity of about 20\%. 
Having established the accuracy of our theory, we proceed to analyze the calculations for TaFeSb with $\text{I}_\text{Fe}$ and $\text{Sb}_\text{Ta}$ defects. We considered two different concentrations, namely 3.7\% and 12.5\%, which provide us with an overview of what may happen when going farther from the dilute limit. 
The presence of defects leads to a significant reduction of the lattice thermal conductivity, which is more marked for $\text{I}_\text{Fe}$ than for $\text{Sb}_\text{Ta}$. This is likely due to the presence of a flat phonon band arising from the interstitial Fe impurity, located just at the top of the acoustic branches, see Figure~\ref{fig:pdos}. Overall, incorporating defects makes the theoretical curves to be much closer to the experimental ones, especially in the region between 400~K and 800~K.
This region is the one that is most important for assessing the validity of our microscopic description. At lower temperatures, the RTA becomes less adequate, because normal scattering dominates, while umklapp scattering is rare~\cite{PhysRevB.104.245424,PAOFLOW}. At higher temperatures, other effects become relevant, as e.g. the grain boundary resistance in polycristalline samples~\cite{GB-HH,GB-TMO} or polaron hopping charge transfer~\cite{POLHOP1,POLHOP2}, which is particularly important for transition metal oxides. Considering that the MTP+MD has shown to overestimate the data of pure TaFeSb, Figure~\ref{fig:k_lat} suggests that a high concentration of defects is unlikely, and probably the system is closer to the dilute limit, when possibly a variety of defects appear together in the crystal. This is in line with the fact that the formation energies presented in Figure~\ref{fig:delta_e} are all above 1 eV per defect.

\section{Conclusion}\label{conc-sec}
In this study we investigated the impact of intrinsic defects on the electronic, phononic, and thermal transport properties of the thermoelectric HH compound TaFeSb, by using a combination of DFT, DFPT, MLIPs, and BTE in the RTA. Through the calculation of the formation energies, it was identified that the most likely defects to arise during growth are Fe additional atoms occupying the vacant lattice in the HH structure ($\text{I}_\text{Fe}$) and Sb substitution at the Ta site ($\text{Sb}_\text{Ta}$). These defects introduce localized states within the electronic structure, reducing the theoretical band gap from the pristine case, and bringing it closer to the experimentally observed values, thereby bridging a notable gap between theoretical predictions and empirical data. The localization of these impurity-like states depends substantially on the nature of the defects, leading to a wider, more itinerant, electronic cloud for $\text{Sb}_\text{Ta}$ than for $\text{I}_\text{Fe}$.
Phonon calculations demonstrated that these defects generate localized phonon modes within the acoustic-optical phonon gap of pure TaFeSb, contributing significantly to the reduction in lattice thermal conductivity ($\kappa_L$). This reduction is more pronounced for $\text{I}_\text{Fe}$ than for $\text{Sb}_\text{Ta}$, effectively enhancing the thermoelectric performance by minimizing heat conduction through phonons. The localized phonon modes are mostly dominated by states deriving from the additional Fe for the $\text{I}_\text{Fe}$ case and by those deriving from the Fe atoms in the neighborhood of the substituted Sb atom for the $\text{Sb}_\text{Ta}$ case. These considerations are in line with the localization of the impurity states discussed for the electronic structure. A detailed comparison of calculated $\kappa_L$ values against experimental measurements confirmed that accounting for these intrinsic defects provide more realistic results, in particular for the temperature range of 400-800 K. Lower and higher temperature can still be improved through the usage of more advanced modeling of transport and grain boundaries.

In Summary, our study underscores the critical role of intrinsic defects in aligning theoretical and experimental values for electronic and thermal properties in HH compounds, which is essential for optimizing their thermoelectric efficiency. Additionally, employing MLIPs has proven to be a powerful tool for exploring defect-induced phonon and thermal transport properties with a high efficiency, offering a viable alternative to the conventional DFT+DFTP method. This approach opens up new avenues for the computational exploration of materials with various types of defects, for varying concentrations, paving the way for enhanced design and predictive capabilities in the research on thermoelectricity.

\section*{Acknowledgments}
We would like to thank D. Nafday and W. Sun for insightful discussions. Most computations were carried out using the computers of the Centre of Informatics Tricity Academic Supercomputer \& Network, Poland. Additional calculations were performed on resources provided by the 
 National Academic Infrastructure for Supercomputing in Sweden (NAISS) and the Swedish National Infrastructure for Computing (SNIC) at National Supercomputer Center (NSC) in Link\"oping, Sweden, partially funded by the Swedish Research Council through Grants No. 2022-06725 and No. 2018-05973. Test-runs were also performed on the local computing clusters (TC and Prime) at the Institute of Physics, Nicolaus Copernicus University, and we are grateful to Szymon {{\'S}miga} for access and support. 
 M.Y.K. and S.M.V.A. are grateful for the financial support from the Iran National Science Foundation (INSF) under Grant No. 99002221. I.D.M. also acknowledges financial support from the European Research Council (ERC), Synergy Grant FASTCORR, Project No. 854843.

%\bibliographystyle{aipnum4}
%\bibliography{Refs}


\end{document}